\documentclass[twocolumn]{aastex631}

\usepackage{graphicx}	% Including figure files
\usepackage{amsmath}	% Advanced maths commands
\usepackage{subfigure}

\begin{document}

\title{Temporal evolution of quasi-periodic oscillations in an accreting black hole Swift J1727.8-1613: coevolution of the disk-corona during the state transition}

\author[0009-0007-6828-3931]{Sai-En Xu}
\affiliation{Department of Astronomy, School of Physics and Technology, Wuhan University, Wuhan 430072, People’s Republic of China}

\correspondingauthor{Bei You}

\author[0000-0002-8231-063X]{Bei You}
\affiliation{Department of Astronomy, School of Physics and Technology, Wuhan University, Wuhan 430072, People’s Republic of China}
\email{youbei@whu.edu.cn}

\author[0009-0007-1966-181X]{Yi Long}
\affiliation{Department of Astronomy, Nanjing University, 163 Xianlin Avenue, Nanjing 210023, People's Republic of China}

\author[0009-0007-7292-8392]{Han He}
\affiliation{Department of Astronomy, School of Physics and Technology, Wuhan University, Wuhan 430072, People’s Republic of China}

\newcommand{\mo}[1]{\color{orange}{#1}}

\begin{abstract}

Low-frequency quasi-periodic oscillations (QPOs) are commonly observed in black hole X-ray binaries, and their frequency has been found to correlate with various spectral properties. In this work, we present a detailed timing analysis of Swift J1727.8-1613, revealing a novel two-branch correlation between the QPO frequency and the observed disk emission, which differs from previous findings of a single correlation. Specifically, at QPO frequencies below 3 Hz, the QPO frequency is negatively correlated with the observed disk emission. This negative relation transitions to a positive one, as the QPO frequency exceeds approximately 3 Hz. The correlation between QPO frequency and Compton flux exhibits an opposite trend, with a positive correlation at lower frequencies and a negative correlation at higher ones. We interpret these behaviors as signatures of an evolving disk-corona geometry, within the framework of a Lense-Thirring precessing hot flow. Additionally, we find that during the flare state, the QPO fractional root-mean-square (rms) remains nearly constant above 15 keV, but increases with energy below this threshold. The slope of the rms-energy relation increases as the energy spectrum softens.

\end{abstract}


\section{Introduction}
\label{sec:intro}

Black hole X-ray binaries (BHXRBs) spend most of their lives in a quiescent state. However, during outbursts, these systems experience dramatic increases in brightness across multi-wavelengths, ranging from radio to gamma-ray emissions \citep{2007A&ARv..15....1D}. The X-ray luminosity during these outbursts can exceed the quiescent level by a factor of more than $10^4$. Based on their energy spectral and timing properties, outbursts can be classified into four distinct states: the low hard state (LHS), the hard-intermediate state (HIMS), the soft-intermediate state (SIMS), and the high soft state (HSS) \citep{2016ASSL..440...61B}.

In the LHS, the X-ray spectrum is primarily dominated by a non-thermal power-law component, which is attributed to the Comptonization of disk photons in a hot, optically thin flow, also known as corona \citep{2021NatCo..12.1025Y}. In contrast, during the HSS, the dominant emission shifts to thermal radiation from a geometrically thin disk. The HIMS and SIMS lie between the LHS and HSS and can be further distinguished by their timing properties and multi-wavelength characteristics \citep{2004PThPS.155...99Z, 2022hxga.book....9K}.
In the truncated disk model \citep[e.g.,][]{1997ApJ...489..865E}, the thin accretion disk is truncated outside the innermost stable circular orbit (ISCO) during the LHS, with an inner hot flow existing between the truncation radius and the ISCO. As the thin disk moves inward toward the black hole with the decrease in the truncation radius, the energy spectrum becomes softer, and the system gradually transitions from the hard state to the soft state and vice versa \citep{you2023Sci}.

The timing properties of BHXRBs are typically investigated using power density spectra (PDS). In the outburst phase, PDS are typically characterized by broadband noise, accompanied by a narrow peak with a centroid frequency below $\sim30\,\text{Hz}$, referred to as the low-frequency quasi-periodic oscillations (QPOs) \citep{2016ASSL..440...61B}. In addition, the QPOs can be further classified into type A, B, and C based on their features in PDS, e.g. centroid frequency $\nu$, quality factor $Q = \nu/(2\Delta)$, where $\Delta$ is the half width at half maximum (HWHM), and the fractional root mean square (rms) \citep{1999ApJ...526L..33W, 2001ApJS..132..377H, 2002ApJ...564..962R,2023MNRAS.520.5544G,2023MNRAS.525..854M,2023MNRAS.526.3944Z,2023MNRAS.520.5144Z,2024ApJ...969..152W}. Characterized by flat-topped noise, high fractional rms ($\sim 20\%$) and a narrow peak ($Q\gtrsim 5$), the commonly observed Type-C QPOs typically appear in LHS and HIMS \citep{2016ASSL..440...61B}.

The broadband noise was proposed to originate from fluctuations propagating through the accretion flow \citep{1997MNRAS.292..679L, 2009MNRAS.397..666W}. Recently, a Monte Carlo simulation of fast variabilities, which includes propagating fluctuations and reverberation mapping, was implemented by \cite{2025ApJ...985..258Z}. In their work, the simulations were applied to successfully explain the observations of MAXI J1820+070 \citep{2019Natur.565..198K,2021A&A...654A..14D}. However, the origin of the QPOs is a matter of debate. Numerous models have been proposed to explain the origin of QPOs, particularly Type-C QPOs. These models can generally be categorized into two classes: geometric and intrinsic \citep{Ingram2019NewAR}. Geometric models attribute the origin of QPOs to variations in the projected geometry along the line of sight, such as the Lense-Thirring precession of the inner hot flow \citep{Ingram2009MNRAS.397L.101I,2018ApJ...858...82Y,2020ApJ...897...27Y} or the precessing jet \citep{2021ApJ...919...92B,2021NatAs...5...94M}. In contrast, intrinsic models suggest that QPOs arise from inherent instabilities, such as oscillations in accretion rate or pressure \citep[e.g.][]{2010MNRAS.404..738C, 2020MNRAS.492.1399K, 2022MNRAS.515.2099B}. In recent years, several observational results, including the inclination dependence of QPO fractional rms amplitude \citep{2015MNRAS.447.2059M} and time lags \citep{2017MNRAS.464.2643V}, have been reported, which support the geometric interpretation. Additionally, the absence of a disk component in the QPO fractional rms spectrum indicates that the disk emission is not modulated \citep{2006MNRAS.370..405S}. However, observational studies have also revealed a tight positive correlation between QPO frequency and disk luminosity, implying that the QPO frequency may be regulated by the mass accretion rate through the thin disk \citep[e.g.][]{1999ApJ...527..321M, Remillard2006ARA&A..44...49R,  Shui2021MNRAS.508..287S}. These observational results highlight that the origin and physical mechanism of QPOs remain unclear. Thus, further exploring the correlations between QPO properties and spectral parameters is important to advance our understanding of QPOs.

Swift J1727.8-1613, a new BHXRB, was first discovered by the Swift/BAT on August 24, 2023 \citep{Negoro2023ATel, wang2023ATel16209}. Following this discovery, several instruments, including {\it Insight}-HXMT, monitored the source and detected a rapid increase in X-ray emissions.
During the initial phase of the outburst, strong Type-C QPOs were observed \citep{Mereminskiy2023arXiv}. Subsequently, the QPO frequency increased dramatically from $80\,\text{mHz}$ on MJD 60181.46 to around $1.2\,\text{Hz}$ after a week, eventually reaching approximately $10\,\text{Hz}$ \citep{Katoch2023ATel16235, Katoch2023ATel16243, Mereminskiy2023arXiv}. This increase in frequency was associated with the evolution of the source from the Low Hard State (LHS) to the High Intermediate State (HIMS) and ultimately to the Soft Intermediate State (SIMS) \citep{Draghis2023ATel16219, 2025MNRAS.540.1394B,2024ApJ...968..106Z}. The X-ray polarimetric evolution was tracked by IXPE \citep{2023ApJ...958L..16V, 2024ApJ...968...76I}. Recently, the evolution of X-ray polarization with the QPO phase was also studied \citep{2024ApJ...961L..42Z}.

In this work, we performed a timing analysis using the observations carried out by {\it Insight}-HXMT and then investigated the correlation between the QPO frequency and the energy spectral fitting parameters. The observations and data reduction are presented in Section \ref{sec:data}. The results of the timing analysis are presented in Section \ref{sec:results}. In Section \ref{sec:discussion}, we present the correlations between QPOs and spectral fitting parameters. A brief conclusion is then given in Section \ref{sec:conclusion}.

\section{Observations and data reduction} 
\label{sec:data}

The Hard X-ray Modulation Telescope ({\it Insight}-HXMT) comprises three instruments: Low Energy X-ray Telescope (LE, $1$-$15\,\text{keV}$), Medium Energy X-ray Telescope (ME, $5$-$30\,\text{keV}$), and High Energy X-ray Telescope (HE, $20$-$250\,\text{keV}$). The follow-up observations monitored by {\it Insight}-HXMT were triggered on August 25, 2023, following the discovery of Swift J1727.8-1613 by MAXI and Swift. During the outburst, {\it Insight}-HXMT performed follow-up monitoring at a high cadence of less than one day \citep{2024ApJ...973...59S}. 

We used the {\it Insight}-HXMT Data Analysis software ({\tt HXMTDAS, v2.06})\footnote{http://hxmten.ihep.ac.cn/software.jhtml} for the data reduction and filtered the data using the default criteria recommended by the {\it Insight}-HXMT team\footnote{http://hxmtweb.ihep.ac.cn/SoftDoc/847.jhtml}. According to the timing analysis of Swift J1727.8-1613, the rising hard state of the outburst lasted from August MJD 60180 to 60189 \citep{he2025dailyfluctuationspropagatedamply}. Then the source remained in the state transition for nearly 43 days until MJD 60233, when it entered the soft state. During this long state transition, a sequence of apparent flares was detected in LE count rates, and these flares were attributed to the variable disk emission, whereas the high-energy variations were much more insignificant \citep{Yu2024MNRAS.529.4624Y, Cao2025arXiv250305411C, he2025dailyfluctuationspropagatedamply}. Based on this flaring phase, the outburst can be categorized into two states: the normal state, in which the source exhibits similar behavior to that of typical BHs, and the flare state characterized by the observed multiple LE flares \citep{Yu2024MNRAS.529.4624Y, he2025dailyfluctuationspropagatedamply}.

Additionally, the good time intervals (GTIs) of ME span longer durations than those of LE and HE. We thereby separated the observational data from three instruments based on ME GTIs. According to the calibration tests since its launch, the data in $2$-$10\,\text{keV}$ for LE, $10$-$30\,\text{keV}$ for ME, and $30$-$100\,\text{keV}$ for HE were used for analysis, with a systematic error of 1.5\% added for three instruments \citep{2021NatCo..12.1025Y}. 

\section{Results}
\label{sec:results}
\subsection{Spectral analysis}
\label{subsec:spectral analysis}

We performed spectral fitting to the observations by {\it Insight}-HXMT in the LE, ME, and HE bands within the $2$-$100\,\text{keV}$ energy range using {\tt XSPEC}, version 12.13.0\footnote{https://heasarc.gsfc.nasa.gov/xanadu/xspec/} \citep{1996ASPC..101...17A}. We excluded the data in the $21$-$24\,\text{keV}$ range due to electron photoelectric effects in the Silver K-shell of the ME detector.

Our analysis showed that the spectra were well fitted using the model {\tt constant*tbabs*(thcomp$\otimes$diskbb+\\relxillCp)}. In this model, the {\tt tbabs} component accounts for Galactic absorption, with a fixed column density of \( N_{\rm H} = 0.226 \times 10^{22}\,\rm cm^{-2} \) \citep{OConnor2023ATel}. The {\tt diskbb} model represents the thermal emission from the accretion disk, while the {\tt thcomp} model describes the Comptonized emission from thermal hot electrons \citep{Zdziarski2020MNRAS.492.5234Z}. The {\tt relxillCp} model is utilized for relativistic reflection, incorporating both Comptonization and reflection on the disk \citep{Garcia2014ApJ...782...76G}.

During spectral fitting, the photon index $\Gamma$ and the electron temperature $kT_{\rm e}$ in {\tt relxillCp} were tied to the corresponding values in {\tt thcomp}. We fixed the iron abundance at \( A_{\rm Fe}=1.0 \) in solar units and set the inclination angle at \( \theta=40^\circ \) \citep{peng2024ApJ...960L..17P}. Furthermore, we maintained the reflection fraction at \( R_{\rm f}=-1 \) to isolate the reflected component \citep{Dauser2016A&A...590A..76D}. The input inner radius in the {\tt relxillCp} model was normalized to the {\tt diskbb} component (see Eq. \ref{eq:rin} in Sec. \ref{subsec:correlations}), while the outer radius was fixed at \( R_{\rm out}=1000R_{\rm g} \), where \( R_{\rm g}=GM/c^2 \) denotes the gravitational radius. For simplicity, the index $q_{1}$ and ${q_2}$ of the emissivity profiles were fixed at 3.0. 
The density of the accretion disk was fixed at $10^{15}\,\text{cm}^{-3}$, a commonly adopted value in spectral fitting. However, we note that recent studies suggest the disk density in BHXRBs can reach values as high as $10^{20}\,\text{cm}^{-3}$ \citep[e.g.,][]{2021ApJ...909..146C}. In addition, acceptable fits can still be obtained even when the assumed disk density is biased \citep{2024ApJ...974..280D, 2025arXiv250600623Z}.
The temperature at inner disk radius $T_{\rm in}$, the covering fraction $c_f$, the ionization of the accretion disk $\log{\xi}$ and the normalization of {\tt relxillCp} were left free. We obtained good fits with reduced $\chi^2 < 1.2$. In the Appendix, we present representative spectra from Exposure IDs P061433801402, P061433802003 and P061433802703 (corresponding to MJD 60200.2, MJD 60206.3, and MJD 60213.3), together with their best-fitting models in Figure~\ref{fig:spectral_fitting_all}.

% \begin{figure}[htb]
%     \centering
%     \includegraphics[width=0.48\textwidth]{P061433802003_1_eeuf_cttdr.pdf}
%     \caption{ Upper panel: Unfolded spectrum obtained from the first GTI of Exposure ID P061433802003. The LE, ME, and HE data are colored in green, blue, and red, respectively. The black solid line denotes the best fit model, which is decomposed into the convolution component {\tt thcomp$\otimes$diskbb}, (purple), and the reflection component {\tt relxillCp} (sky blue). Lower panel: Data-to-model ratio.}
%     \label{fig:spectral_fitting}
% \end{figure}

The luminosities in 0.01-1000 keV of the disk and the Compton components were estimated using {\tt cflux} in {\tt XSPEC}. The {\tt diskbb} flux $F_{\rm disk}$ corresponds to the intrinsic disk emission, which comprises the seed photons for Compton cooling and the unscattered soft photons that escape the accretion flow (i.e., the observed emission). The {\tt thcomp} component incorporates the flux of the Comptonized photons plus the unscattered soft photons from the disk \citep{Zdziarski2020MNRAS.492.5234Z}. Therefore, the unscattered flux $F_{\text{us-disk}}$ directly observed from the accretion disk was derived through $F_{\text{us-disk}} = (1-c_{f})F_{\rm disk}$, and the Compton flux was derived through $F_{\rm Comp}=F_{\rm thcomp}-(1-c_f)F_{\rm disk}$, where $c_f$ is the covering fraction \citep{Zdziarski2020MNRAS.492.5234Z}. The reflection flux was obtained from the flux of the {\tt relxillCp} component. The corresponding luminosities were calculated assuming a distance of $d=3.7\,\text{kpc}$ \citep{Mata2025A&A...693A.129M}. We include the luminosity and some key fitting parameters in the Appendix, which are taken from \citet{he2025dailyfluctuationspropagatedamply}.

\subsection{Timing analysis}
\label{subsec:timing analysis}

The low-frequency QPOs were detected by three instruments during the outburst. To investigate the properties of these QPOs, we utilized the {\tt powspec} package in {\tt HEASoft} version 6.31\footnote{https://heasarc.gsfc.nasa.gov/docs/software/lheasoft/} to compute the PDS in the LE, ME, and HE for each observation. In each analysis, we used a time segment of 64 seconds, with a time resolution of 1/128 seconds, resulting in a frequency band from $1/64\,\text{Hz}$ to $64\,\text{Hz}$. The PDS was then normalized using Miyamoto normalization \citep{Miyamoto1991ApJ...383..784M} and rebinned in frequency space with a geometrical factor of 1.03, after subtracting the Poisson noise.

\begin{figure*}[htb]
    \centering
    \includegraphics[width=1\textwidth]{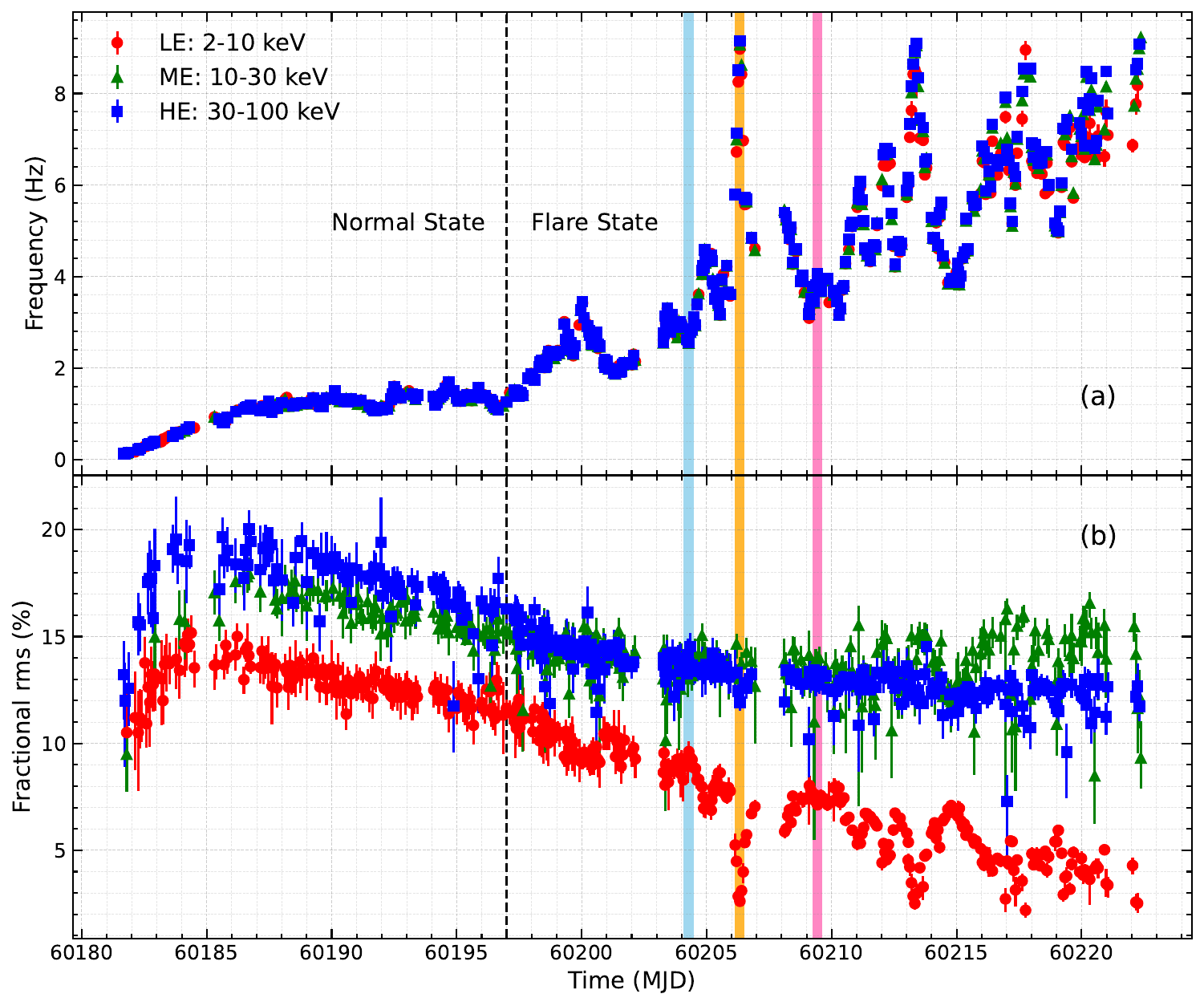}
    \caption{Temporal evolution of the QPO centroid frequency (upper panel) and fractional rms (lower panel). Red circles, green triangles, and blue squares represent data in the LE ($2$-$10\,\text{keV}$), ME ($10$-$30\,\text{keV}$), and HE ($30$-$100\,\text{keV}$) bands, respectively. The transition from the normal state to the flare state is indicated by the dashed vertical line.}
    \label{fig:qpo_LMH}
\end{figure*}

The PDS was then fitted with four {\tt lorentz} models in {\tt XSPEC}, following the methods in \cite{2002ApJ...572..392B}. Considering the contribution from the background, the fractional rms of QPO was corrected using the formula ${\rm rms} = \sqrt{R}\times (S+B)/S$, where $R$ represents the integral of the QPO component in PDS, $S$ denotes the source count rate, and $B$ is the background count rate \citep{1994A&A...292..175B}. The temporal evolutions of the QPO frequency and fractional rms are presented in Fig. \ref{fig:qpo_LMH}.  

We follow the outburst phase classification scheme in \citet{Yu2024MNRAS.529.4624Y}, referring to the phases before and after MJD 60197 as normal state and flare state, respectively. In the normal state, the QPO frequency increases slowly to $1.3\,\text{Hz}$ on MJD 60190, and then remains around $1.3\,\text{Hz}$ until MJD 60197 when the source enters the flare state. During the flare state, the QPO frequency exhibits pronounced variability that resembles the LE light curve, exhibiting multiple peaks with rapid variations. A particularly striking example occurs near MJD 60206, where the frequency undergoes a 125\% increase from $4\,\text{Hz}$ to $9\,\text{Hz}$, followed by a 33\% decrease from $9\,\text{Hz}$ to $6\,\text{Hz}$ within a 1-day window.

The fractional rms in all three instruments rises rapidly during the initial phase of the outburst, reaching a peak around MJD 60185, after which a gradual decline begins. After about MJD 60203, the fractional rms in the LE band continues to decrease from $\sim$10\% to $\sim$3\%, while in the ME and HE bands, it remains stable at around 15\%. Notably, dips in the LE-band fractional rms coincide with peaks in the QPO frequency flares, indicating a weakening of the QPO during these intervals.

\section{Discussion}
\label{sec:discussion}

In the 2023 outburst, Swift J1727.8-1613 exhibits a sequence of flares in the LE band during the flare state, predominantly contributing to the disk variability (see Figure~\ref{fig:flux_cttdr}) driven by the disk instability \citep{he2025dailyfluctuationspropagatedamply}. The reflection component shows an overall decline over time; however, the covering fraction $c_f$, intrinsic disk luminosity, unscattered disk luminosity, and Compton luminosity exhibit more complex patterns, which appear to follow two distinct behaviors with a transition around MJD 60203. Before approximately MJD 60203, $c_f$ reaches a peak around MJD 60200, while the intrinsic disk luminosity remains nearly constant. This leads to a dip in the unscattered disk luminosity and a corresponding peak in the Compton luminosity. After MJD 60203, the intrinsic disk luminosity undergoes a series of flares, which are accompanied by inverse dips in $c_f$. As a result, the unscattered disk luminosity also displays a series of flares, while the Compton luminosity generally follows a gradual decreasing trend.

Our timing analysis also reveals that the QPO frequency varies during this state, exhibiting multiple peaks with rapid changes, which appear to correlate with the luminosity. Specifically, before MJD 60203, the QPO frequency follows a trend similar to that of the Compton emission. However, after MJD 60203, it aligns more closely with the behavior of the disk emission.

In this section, we investigate the correlations between the energy spectral parameters and the QPO frequency, and explore potential scenarios for their physical connection. Additionally, we discuss the energy dependence of fractional rms during the flares.

\subsection{QPO frequency-luminosities correlations}
\label{subsec:correlations}

\begin{figure*}[htbp]
    \centering
    \includegraphics[width=0.95\textwidth]{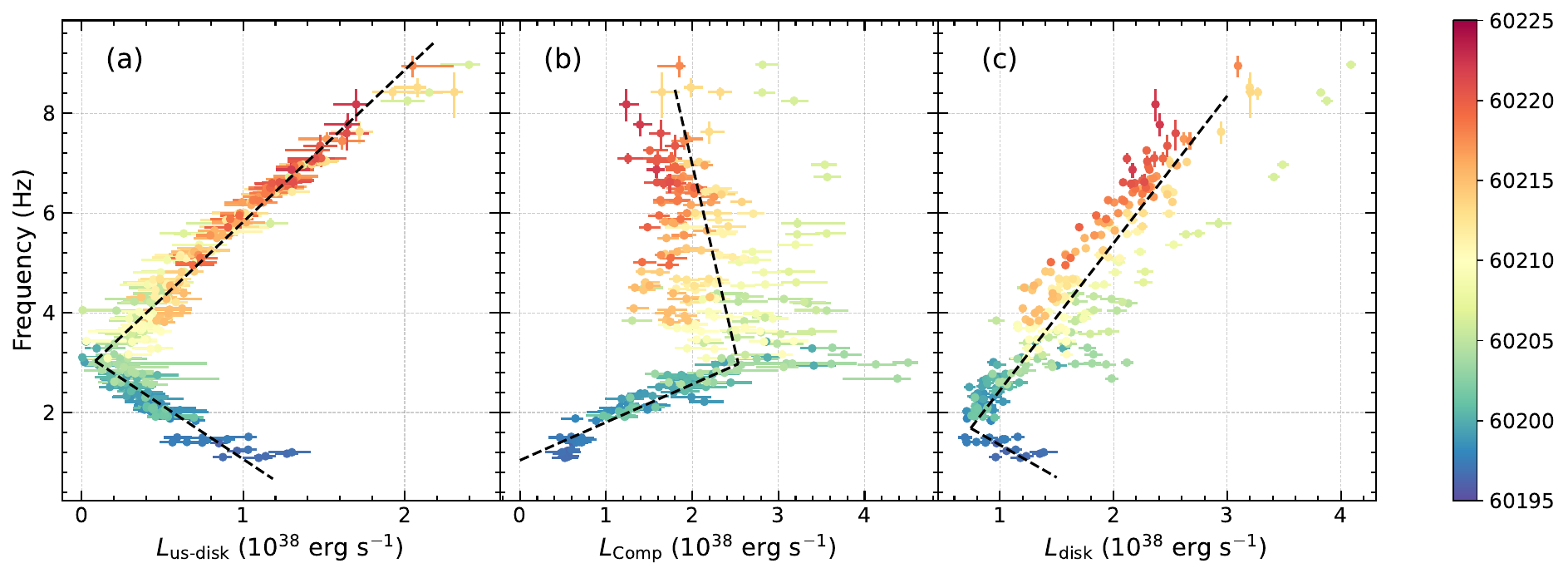}
    \caption{The correlations between QPO frequency and (a) unscattered disk luminosity, (b) Compton luminosity, and (c) intrinsic disk luminosity during flare state. The MJD is mapped to the colors of the points. The dashed curves indicate the best-fit results.
    }
    \label{fig:lefre_flux}
\end{figure*}

Considering that the QPO frequencies are nearly identical across the three energy bands, we adopt the QPO frequency measured in the LE band for the subsequent analysis. In Fig.~\ref{fig:lefre_flux} we present the correlations between QPO frequency $\nu$ and: (a) unscattered disk luminosity $L_{\text{us-disk}}$; (b) Compton luminosity $L_{\rm Comp}$; (c) intrinsic disk luminosity $L_{\rm disk}$. Both the $\nu$–$L_{\text{us-disk}}$ and the $\nu$–$L_{\rm Comp}$ relations display two-branch correlations with a similar transition frequency around $3\,\text{Hz}$. The $\nu$–$L_{\rm disk}$ also exhibits two-branch correlation. Specifically, for the branch of lower frequencies ($\nu < 2\,\text{Hz}$), the QPO frequency appears to be independent of $L_{\rm disk}$ (see below). Instead, at higher frequencies, there is a strong positive correlation between the QPO frequency and $L_{\rm disk}$, which can be well described by a power-law function of the form $\nu \propto L_{\rm disk}^{a}$, with the best-fit exponent $a = 0.967 \pm 0.023$, consistent with a nearly linear scaling.

Therefore, given the distribution described above, we adopt, for simplicity, piecewise linear functions for these three relations. The piecewise linear function is given by:
\begin{equation}
    L =
    \begin{cases}
        f_1 (\nu - \nu_{\rm tr}) + L_{\rm tr}, & \nu < \nu_{\rm tr}, \\
        f_2 (\nu - \nu_{\rm tr}) + L_{\rm tr}, & \nu > \nu_{\rm tr}.
    \end{cases}
    \label{eq:fit}
\end{equation}

The fitting was performed using the Levenberg–Marquardt method implemented in the Python package LMFIT \citep{newville_2014_11813}, treating $\nu$ as the independent variable for convenience. We found the transition frequencies for $\nu$–$L_{\text{us-disk}}$ and $\nu$–$L_{\rm Comp}$ to be $3.03\pm 0.03\,\text{Hz}$ and $2.97\pm 0.09\,\text{Hz}$, respectively. For ease of interpretation, we rewrite the best-fit relations in the form $\nu = aL + b$. For the $\nu$–$L_{\text{us-disk}}$ relation, we obtain:
\begin{equation} 
\begin{aligned}
    \nu = (-2.15\pm 0.10) L_{\text{us-disk}} (/10^{38} {\rm erg\, s^{-1}}) + (3.21\pm 0.04)&,\\ 
    \text{for}\,\nu < 3.03\, \text{Hz}&;\\
    \nu = (3.04\pm 0.05) L_{\text{us-disk}} (/10^{38} {\rm erg\, s^{-1}}) + (2.78 \pm 0.05)&,\\
    \text{for}\,\nu \geq 3.03\, \text{Hz}&.
\end{aligned}
\end{equation}
We used Spearman correlation coefficient with Monte Carlo error analysis \citep{2014arXiv1411.3816C} to discern the correlations. The Spearman coefficients for the data below and above $3.03\,\text{Hz}$ are $r_{\rm S}=-0.794$ and $r_{\rm S}=0.910$ with a significance of $4.4 \, \sigma$ and $8.7 \, \sigma$, respectively. For the $\nu$–$L_{\rm Comp}$ relation, the best-fit results are:
\begin{equation} 
\begin{aligned}
    \nu = (0.76\pm 0.06) L_{\rm Comp} (/10^{38} {\rm erg\, s^{-1}}) + (1.04\pm 0.18)&,\\
    \text{for}\,\nu < 2.97\, \text{Hz}&;\\
    \nu = (-7.48\pm 1.46) L_{\rm Comp} (/10^{38} {\rm erg\, s^{-1}}) + (21.94 \pm 3.73)&,\\
    \text{for}\,\nu \geq 2.97\, \text{Hz}&.
\end{aligned}
\end{equation}
The Spearman coefficients for the data below and above $2.97\,\text{Hz}$ are $r_{\rm S}=0.910$ and $r_{\rm S}=-0.348$ with a significance of $6.0 \, \sigma$ and $2.1 \, \sigma$, respectively. 
 
For the $\nu$–$L_{\text{disk}}$ relation, at low frequencies range ($\nu < 1.68\, {\rm Hz}$), there is a weak correlation with a Spearman coefficient of $r_{\rm S}=-0.371$, corresponding to a significance of $0.7\,\sigma$. In contrast, at high frequencies range ($\nu \geq 1.68\, {\rm Hz}$), the correlation becomes strongly positive, with a Spearman coefficient of $r_{\rm S}=0.914$ and a significance of $10.4\,\sigma$. The best-fitting results are as follows.
\begin{equation} 
\begin{aligned}
    \nu = (-1.30\pm 0.76) L_{\rm disk} (/10^{38} {\rm erg\, s^{-1}}) + (2.66\pm 0.59)&,\\
    \text{for}\,\nu < 1.68\, \text{Hz}&;\\
    \nu = (2.95\pm 0.08) L_{\rm disk} (/10^{38} {\rm erg\, s^{-1}}) + (-0.51 \pm 0.23)&,\\
    \text{for}\,\nu \geq 1.68\, \text{Hz}&.
\end{aligned}
\end{equation}
We note that a similar positive correlation in Swift J1727.8-1613 was also reported in \citet{Cao2025arXiv250305411C}, and the quantitative correlation is provided here. 

A strong positive correlation has been reported between the QPO frequency and the unscattered disk flux (i.e., photons directly observed from the accretion disk) in previous studies. In GRS 1915+105, the thermal disk flux was found to increase with QPO frequency \citep{1999ApJ...513L..37M, 1999ApJ...527..321M, Muno2001ApJ...556..515M, Mikles2006ApJ...637..978M}. Similar correlations between Type-C QPO frequency and unscattered disk flux were confirmed in XTE J1550-564, GRO J1655-40, H1743-322, and GX 339-4 \citep{Sobczak2000ApJ...531..537S, Remillard2002ApJ...564..962R, Rao2010ApJ...714.1065R, Motta2011MNRAS.418.2292M, Shui2021MNRAS.508..287S}. Notably, in GX 339-4, a positive correlation with disk flux was observed for all three types of QPOs: Type-A, B, and C \citep{Motta2011MNRAS.418.2292M}.

In contrast, the correlation between QPO frequency and Compton flux is more complex and appears to depend on both the source and QPO type. A positive correlation has been reported in XTE J1550-564 (Type-C), H1743-322 (Type-C) and GRS 1915+105 \citep{Remillard2006ARA&A..44...49R, 1999ApJ...527..321M}, whereas a negative correlation is observed in GRO J1655-40 \citep{Sobczak2000ApJ...531..537S, 1999ApJ...527..321M}. Moreover, in GX 339-4, the correlation is positive for Type-B QPOs but negative for Type-C \citep{Motta2011MNRAS.418.2292M}.

However, for Swift J1727.8–1613, we observe, for the first time, a negative correlation between Type-C QPO frequency $\nu$ and unscattered disk flux $L_{\text{us-disk}}$. Moreover, the presence of a transition frequency in both the $\nu$–$L_{\text{us-disk}}$ and $\nu$–$L_{\rm Comp}$ relations suggests a change in the QPO mechanism during the outburst, indicating a more complex interaction between the thin disk and the inner hot flow. We further explore this scenario in Section~\ref{sec: coevolution}. 

\subsection{Coevolution of the inner hot flow and thin disk}
\label{sec: coevolution}

To further explore the geometry and radiative coupling between the thin disk and the inner hot flow, we examined the evolution of the disk truncation radius and the covering fraction, defined as the fraction of seed photons from the disk that are Comptonized by the inner hot flow.
The truncation radius of the thin disk $R_{\rm tr}$ was estimated using \citep{Mitsuda1984PASJ...36..741M}: 
\begin{equation}
   R_{\rm tr} = \kappa^2 \xi  \sqrt{N_{\rm diskbb}*D_{10}^{2}/\cos{\theta}},
   \label{eq:rin}
\end{equation}
where $N_{\rm diskbb}$ is the normalization of the model {\tt diskbb}, $D_{10}$ is the distance to the source in units of 10 kpc, and $\theta$ is the inclination of the source. Here we used a distance of 3.7 kpc \citep{Mata2025A&A...693A.129M}, and assumed an inclination of $40^{\circ}$ and an experimental parameter of $M_{\rm BH}=10\,\rm M_{\odot}$. The ratio of color temperature to effective temperature, $\kappa$ was set to 1.7 \citep{1995ApJ...445..780S}. The correction factor $\xi$ is supposed to be 0.41 for a non-spinning BH with a disk extending to the ISCO \citep{10.1093/pasj/50.6.667}. 
Since Swift J1727.8-1613 was suggested to harbor a spin of $a>0$ \citep{peng2024ApJ...960L..17P,Yu2024MNRAS.529.4624Y}, and it was unclear whether the inner disk reached the ISCO or not during the state transition, we therefore set $\xi=1$ to neglect the influence of this factor.

\begin{figure*}[ht]
    \centering
    \includegraphics[width=0.95\textwidth]{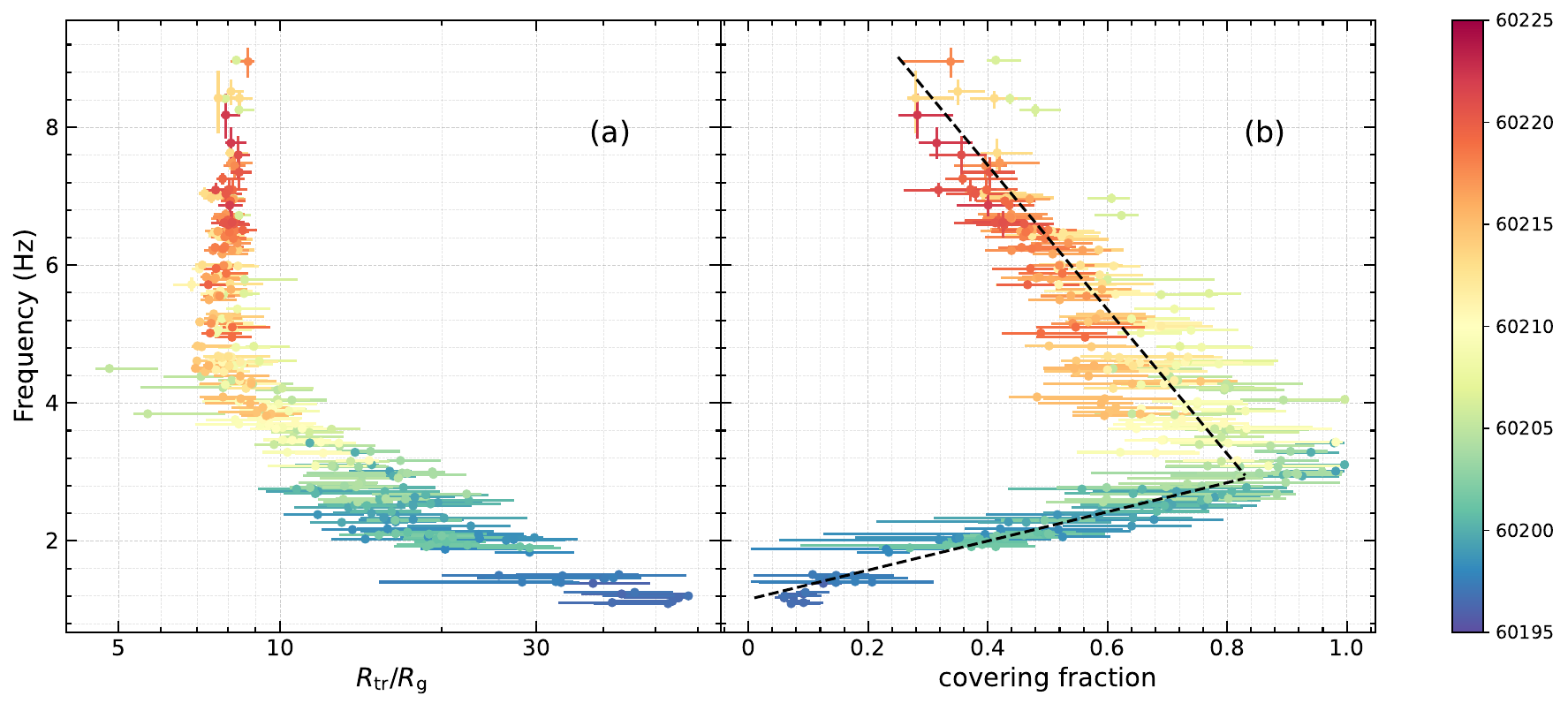}
    \caption{The QPO frequency as a function of (a) disk truncation radius and (b) covering fraction. The radius $R_{\rm tr}$ is normalized by the gravitational radius $R_{\rm g}$. The dashed curves represent the best-fitting results. The MJD is mapped to the colors of the points. }
    \label{fig:lefre_rin_cov}
\end{figure*}

The correlation between the QPO frequency and the disk truncation radius is presented in Fig. \ref{fig:lefre_rin_cov} (a). When the QPO frequency is relatively low ($\nu \lesssim 4.5$ Hz), it shows a negative correlation. However, for higher QPO frequencies ($\nu \gtrsim 4.5$ Hz), the QPO frequency exhibits a rapid variation while the disk truncation radius remains approximately constant.

In the Lense-Thirring precession model, the entire inner hot flow precesses as a solid body. The QPO frequency is set by the inner radius $R_{\rm i}$ (typically at the ISCO) and outer radius $R_{\rm o}$ of the inner hot flow \citep{Ingram2009MNRAS.397L.101I}. As a result, one predicts a negative correlation between QPO frequency and $R_{\rm o}$. Instead, we observe that the QPO frequency changes dramatically while the disk truncation radius remains constant.

Having established how $\nu$ varies with $R_{\mathrm{tr}}$, we now turn to the radiative coupling between the thin disk and the inner hot flow. The QPO frequency $\nu$ as a function of covering fraction $c_f$ is presented in Fig.~\ref{fig:lefre_rin_cov}(b). It displays a distinct two-banch correlation with a transition frequency around $3\,\text{Hz}$. By applying the piecewise linear fitting procedure in Section~\ref{subsec:correlations}, we derive a transition frequency of $2.92\pm0.03\,\text{Hz}$. The best-fit results are:
\begin{equation} 
\begin{aligned}
    \nu = (2.12\pm 0.08) c_f + (1.15\pm 0.08)&,\\
    \text{for}\,\nu < 2.92\, \text{Hz}&;\\
    \nu = (-10.47\pm 0.44) c_f + (11.64 \pm 0.38)&,\\
    \text{for}\,\nu \geq 2.92\, \text{Hz}&.
\end{aligned}
\end{equation}
The Spearman coefficients for the data below and above $2.92\,\text{Hz}$ are $r_{\rm S}=0.854$ and $r_{\rm S}=-0.778$ in a significance of $6.0 \, \sigma$ and $4.9 \, \sigma$, respectively. 

\begin{figure}[t]
    \centering
    \includegraphics[width=0.45\textwidth]{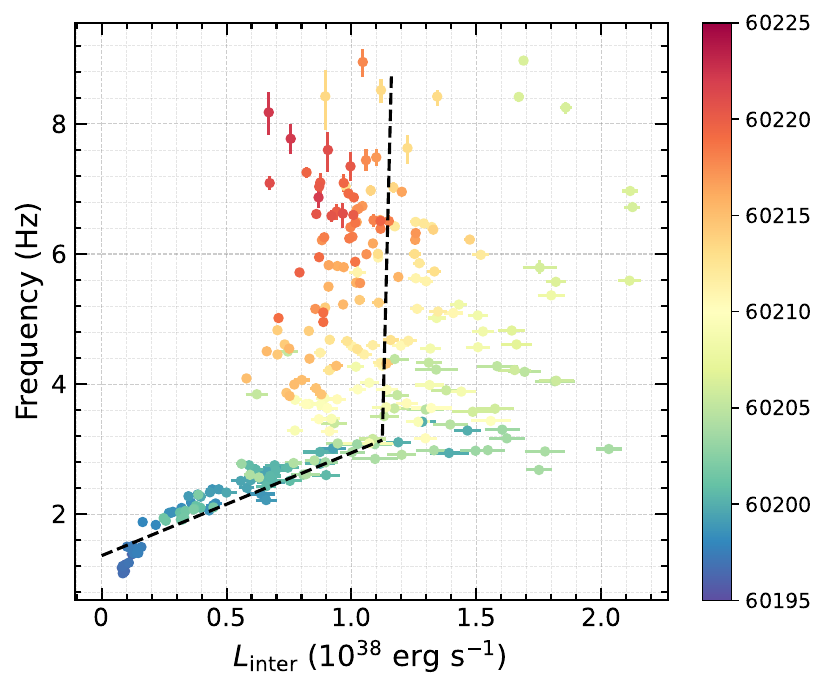}
    \caption{The correlations between QPO frequency and intercepted disk luminosity during flare state. The MJD is mapped to the colors of the points. The dashed curve indicates the best-fitting result.}
    \label{fig: intercepted}
\end{figure}

The fact that the $\nu$-$c_f$, $\nu$-$L_{\text{us-disk}}$, and $\nu$-$L_{\rm Comp}$ correlations share very similar transition frequencies implies that the covering fraction plays a key role in driving the observed anomalous correlations. Since the inner hot flow’s emission is dominated by Compton up‑scattering of the intercepted photons from the thin disk, the emission of inner hot flow $L_{\rm Comp}$ should be closely tied to the intercepted disk luminosity $L_{\rm inter} = c_f L_{\rm disk}$. In Fig.~\ref{fig: intercepted}, we present the $\nu - L_{\rm inter}$ relation, which closely resembles to the $\nu$-$L_{\rm Comp}$ relation, with a simlar transition frequency of $3.14\pm0.10\,\text{Hz}$. This indicates that the $\nu$-$L_{\rm Comp}$ relation may simply reflect the combined effects of $L_{\rm disk}$ and $c_f$.

The above radiative correlation may further reflect changes in the underlying geometry of the thin disk and inner hot flow. In the {\tt thcomp} model \citep{Zdziarski2020MNRAS.492.5234Z}, the covering fraction denotes the ratio of the seed photons to be Comptonized to the intrinsic photons from the disk. For a thin disk with a constant truncation radius, changes in covering fraction can reflect variations in the outer radius $R_{\rm o}$ of the inner hot flow. Specifically, an expansion of the inner hot flow results in a higher fraction of disk photons being Comptonized, corresponding to a larger covering fraction. 

\begin{figure}[tbp]
    \centering
    \includegraphics[width=0.45\textwidth]{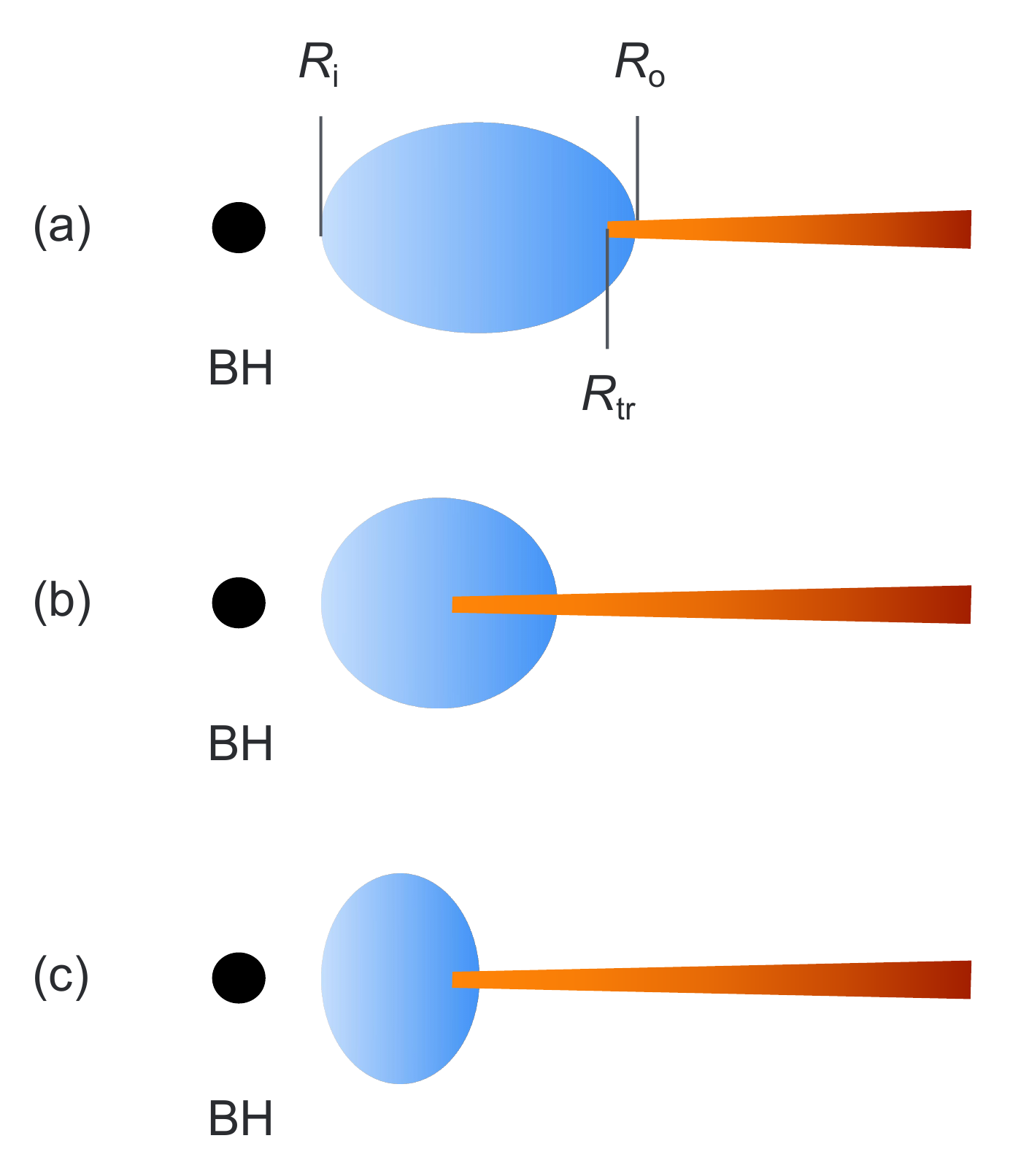}
    \caption{Schematic of the potential coevolution geometry between the inner hot flow and the accretion disk. For clarity, only half of the accretion flow is depicted. The black hole is marked as a solid black circle, the inner hot flow is shown as a blue ellipse, and the accretion disk is colored brown.}
    \label{fig: geometry}
\end{figure}

Based on the above analysis, we propose a possible scenario to interpret the correlation between QPO frequency and other spectral fitting parameters. Fig.~\ref{fig: geometry} shows a toy model of the geometric evolution of the thin disk and inner hot flow. The thin disk is truncated at a radius $R_{\rm tr}$, and the inner hot flow extends between an inner radius $R_{\rm i}$ and an outer radius $R_{\rm o}$.

When the QPO frequency is relatively low, it tends to increase as the truncation radius $R_{\rm tr}$ move inward to the black hole and the covering fraction rises. This corresponds to the evolution from state (a) to state (b) in Fig.~\ref{fig: geometry}. The Compton luminosity $L_{\rm Comp}$ also rises due to an increase in the number of seed photons available for Compton cooling.
At the same time, the outer thin disk extends further inward into the inner hot flow, resulting in a higher covering fraction. This evolution results in a reduction in the unscattered disk luminosity $L_{\text{us-disk}}$ due to increased Compton scattering of soft photons. Moreover, the decrease in the characteristic radius $R_{\rm o}$ contributes to the observed increase in QPO frequency. 

At relatively high QPO frequencies, Fig.~\ref{fig:lefre_rin_cov} (a) shows that $R_{\rm tr}$ remains largely constant. However, the rate of mass accretion through the inner thin disk changes, which is reflected in the changes in the disk emission \citep{he2025dailyfluctuationspropagatedamply}. Enhanced Compton cooling may cause the inner hot flow to contract, resulting in a decrease in $R_{\rm o}$, which in turn contributes to the rise in the QPO frequency. Although the number of seed photons increases, the decrease in covering fraction counteracts this effect on the Compton luminosity $L_{\rm Comp}$. As a result, $L_{\rm Comp}$ becomes more variable and shows a weak correlation with QPO frequency.
In Fig.~\ref{fig: geometry}, the evolution from (b) to (c) illustrates this process. 

We note that the correlations between the QPO frequency and both the unscattered disk luminosity and the Compton luminosity are derived from spectral fits using the convolution model {\tt thcomp$\otimes$diskbb+relxillCp}. 
In Figure~\ref{fig:cttdr_para}, we present the time-evolutions of the fitting parameters of {\tt thcomp} and {\tt relxillCp}. The normalization of reflection, which corresponds to the reflection luminosity, shows a gradual decline over time, while the covering fraction and the luminosity of other components follow two distinct patterns, with a transition occurring around MJD 60203 (see Figures~\ref{fig:flux_cttdr} and \ref{fig:cttdr_para}). This suggests that the observed variation in the covering fraction is not due to the degeneracy between the reflection and convolution components.

Furthermore, we also examine the influence of the chosen spectral model on the derived correlations. To do this, we repeat the analysis with an alternative spectral model, {\tt diskbb+xillver}, which is also widely used in the literature. The evolutionary trends of the reflection, unscattered disk, and Compton luminosities remain broadly consistent between the two models. More importantly, the correlations between QPO frequency and luminosity (Figure~\ref{fig:lefre_flux}) still hold true, reinforcing the robustness of our conclusions. A detailed analysis of the {\tt diskbb+xillver} model is provided in Appendix~\ref{sec:appendix2}.

\subsection{The energy dependence of fractional rms}

\begin{figure}[tbp]
    \centering
    \includegraphics[width=0.45\textwidth]{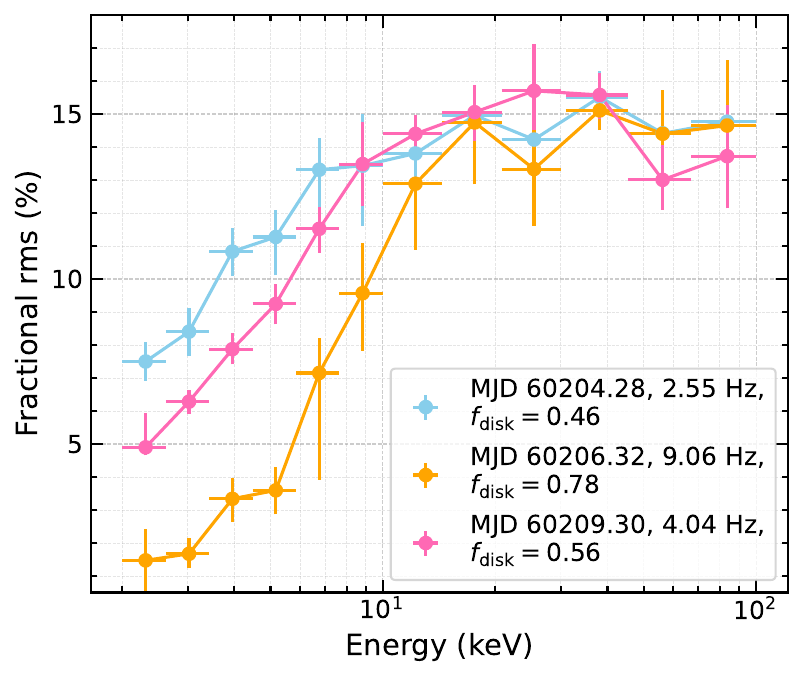}
    \caption{The fractional rms as a function of energy. The three observations are colored with skyblue, orange, and pink, respectively, corresponding to the highlighted regions in Fig.~\ref{fig:qpo_LMH}. Labels indicate MJD, QPO frequency (Hz), and disk fraction $f_{\rm disk}$.} 
    \label{fig:flare_rms}
\end{figure}

As shown in Fig.~\ref{fig:qpo_LMH}, the QPO fractional rms in LE decreases with increasing QPO frequency, while the fractional rms in the ME and HE remain roughly constant at about 15\%, which is consistent with previous findings in EXO 1846–031 reported by \citet{2021RAA....21...70L}. This motivates us to further explore the energy-dependent evolution of QPO fractional rms. The data corresponding to the peaks and dips of the QPO frequency variations, highlighted in Figure \ref{fig:qpo_LMH}, were selected for the following analysis. Then we generated a series of power density spectra across logarithmically spaced energy bands: LE from 2-2.6 keV, 2.6-3.4 keV, 3.4-4.5 keV, 4.5-5.8 keV, 5.8-7.6 keV, 7.6-10 keV; ME from 10-14.4 keV, 14.4-20.8 keV, 20.8-30 keV; and HE from 30-44.8 keV, 44.8-66.9 keV, 66.9-100 keV.

By fitting the PDS in each energy band, we obtained the corresponding QPO fractional rms. Due to the low signal-to-noise ratio in the noise-dominated frequency bands, we fixed the centroid frequencies of the noise components during the error estimation process to minimize their impact on the final rms uncertainties. The background was also corrected using the method described in Section~\ref{subsec:timing analysis}. The fractional rms spectra for three observations (Exposure ID: P061433801801, P061433802003, and P061433802303) are presented in Fig.~\ref{fig:flare_rms} and are labeled with their MJD dates, QPO frequencies, and disk flux fractions $f_{\rm disk} = F_{\rm disk}/F_{\rm thcomp}$. 
Above about $15\,\text{keV}$, the fractional rms remains nearly constant at around 15\%. However, below this energy, the fractional rms increases with energy. Additionally, the slope of the fractional rms spectrum steepens when the QPO frequency is higher and the thermal component is more dominant. An example of this behavior can be seen in the observation at MJD 60206.32, which is plotted in orange in Fig.~\ref{fig:flare_rms}. 

The evolution of the fractional spectra has been explored by \citet{2018ApJ...858...82Y}; they employed a model that combines a truncated disk with a precessing inner hot flow to perform numerical simulations of the X-ray flux variability in BHXRBs during the transition from the hard to the soft spectral state. Their results, Figure 5 in \citet{2018ApJ...858...82Y}, reveal some notable features: (i) Above about $3\,\text{keV}$, where the emission from hot flow dominated over other components, the fractional rms remains approximately constant (in the soft state) or increase with energy (in the hard state); and (ii) there is a steeper slope in the low-energy range (below about $3\,\text{keV}$) when the spectrum is in a softer state, due to the fact that the relatively stable emission of the thin disk becomes dominated at this energy band. The flattening of fractional rms at higher energy band has been observed in some sources \citep[e.g.,][]{2022ApJ...933...69C, 2022MNRAS.512.2686Z, 2023MNRAS.525..854M}.
Our results are also generally consistent with their simulations, although the pivot energy ($15\,\text{keV}$) in our observations is higher. This difference may suggest that the low-energy emission from the hot flow is more stable than predicted when the spectrum becomes softer.

\section{Conclusion}
\label{sec:conclusion}

We have performed a comprehensive timing analysis of the black hole X-ray binary Swift J1727.8$-$1613. Our main findings are summarized as follows:  
\begin{enumerate}  
    \item We report, for the first time, a clear negative correlation between the unscattered disk emission $L_{\text{us-disk}}$ and the QPO frequency $\nu$. This negative relation transitions to a positive one when the QPO frequency is higher than about $3\,\text{Hz}$. The relation between the QPO frequency and Compton emission $L_{\rm Comp}$ also exhibits a two-branch correlation, with a similar transition frequency. But the trend of $\nu$-$L_{\rm Comp}$ is opposite. Specifically, there is a positive correlation at lower frequencies and a negative correlation at higher frequencies.
    \item We interpret these two-branch correlations as the result of a coevolution in the accretion geometry between the outer thin disk and the inner hot flow, within the framework of Lense-Thirring precessing hot flow.  
    \item The energy dependence of the QPO fractional rms during the flare state shows that it remains nearly constant above $15\,\text{keV}$, but increases with energy below this threshold. Furthermore, the slope of the rms–energy relation increases when the energy spectrum softens.
\end{enumerate}

\section*{Acknowledgements}

This work made use of the data from the {\it Insight}-HXMT mission, a project funded by the China National Space Administration (CNSA) and the Chinese Academy of Sciences (CAS), and software provided by the High Energy Astrophysics Science Archive Research Center (HEASARC), a service of the Astrophysics Science Division at NASA/GSFC. All {\it Insight}-HXMT data used in this work are publicly available and can be downloaded from the official website of {\it Insight}-HXMT: http://hxmtweb.ihep.ac.cn.
B.Y. is supported by the National Program on Key Research and Development Project 2021YFA0718500; by NSFC grants 12322307, 12273026, and 12361131579; by Xiaomi Foundation / Xiaomi Young Talents Program.

\appendix

\section{Details of spectral parameters and luminosity decomposition}
\label{sec:appendix1}

\begin{figure}[htb]
    \centering
    \includegraphics[width=0.9\textwidth]{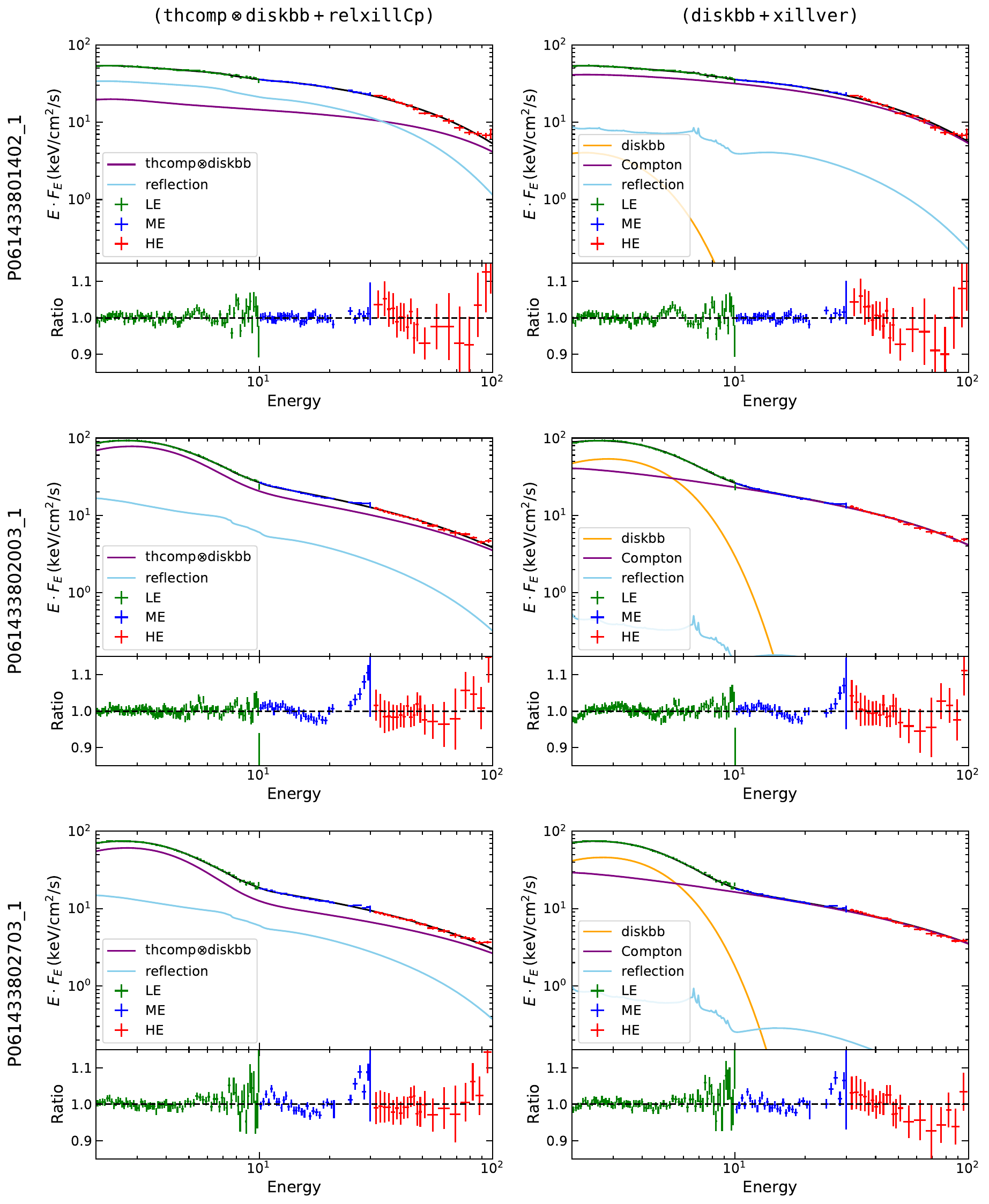}
    \caption{ Unfolded spectra obtained from the first GTI of Exposure ID P061433801402, P061433802003 and P061433802703. Results from the spectral model {\tt (thcomp$\otimes$diskbb+relxillCp)} are shown in the left column, and those from the spectral model {\tt (diskbb+xillver)} in the right column. In the left panels, the convolution component ({\tt thcomp$\otimes$diskbb}) is shown in purple, and the reflection component in sky blue. In the right panels, the diskbb, Compton and reflection components are plotted in orange, purple and sky blue. The LE, ME and HE data are plotted in green, blue and red, respectively. All spectra are rebinned for visual clarity.}
    \label{fig:spectral_fitting_all}
\end{figure}

\begin{figure}[tbp]
    \centering
    \includegraphics[width=0.65\textwidth]{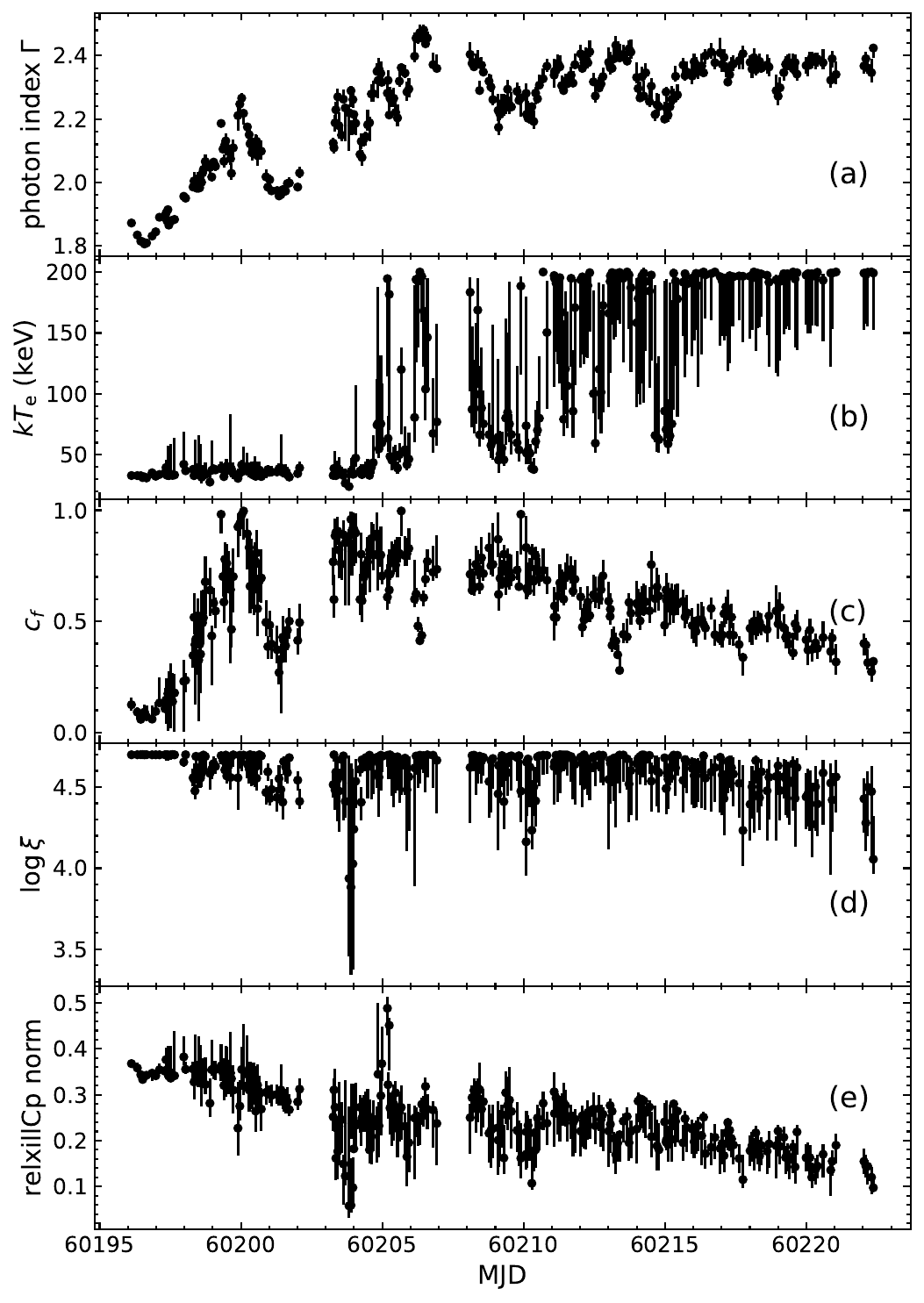}
    \caption{Key fitting parameters of {\tt thcomp} and {\tt relxillCp} which are taken from the spectral fits in \cite{he2025dailyfluctuationspropagatedamply}. (a): The power-law photon index. (b): The electron temperature. (c): The covering fraction. (d): The ionization of the accretion disk. (e): The normalization of {\tt relxillCp}.} 
    \label{fig:cttdr_para}
\end{figure}

In the left column of Figure~\ref{fig:spectral_fitting_all}, we show the unfolded spectra and the component decomposition of the model {\tt constant*tbabs*(thcomp$\otimes$diskbb+relxillCp)}, based on the spectral fits in \cite{he2025dailyfluctuationspropagatedamply}. Key fitting parameters of {\tt thcomp} and {\tt relxillCp} are presented in Figure~\ref{fig:cttdr_para}. During the fitting, the photon index $\Gamma$ and the electron temperature $kT_{\rm e}$ of {\tt thcomp} and {\tt relxillCp} were tied.

\begin{figure}[tbp]
    \centering
    \includegraphics[width=0.65\textwidth]{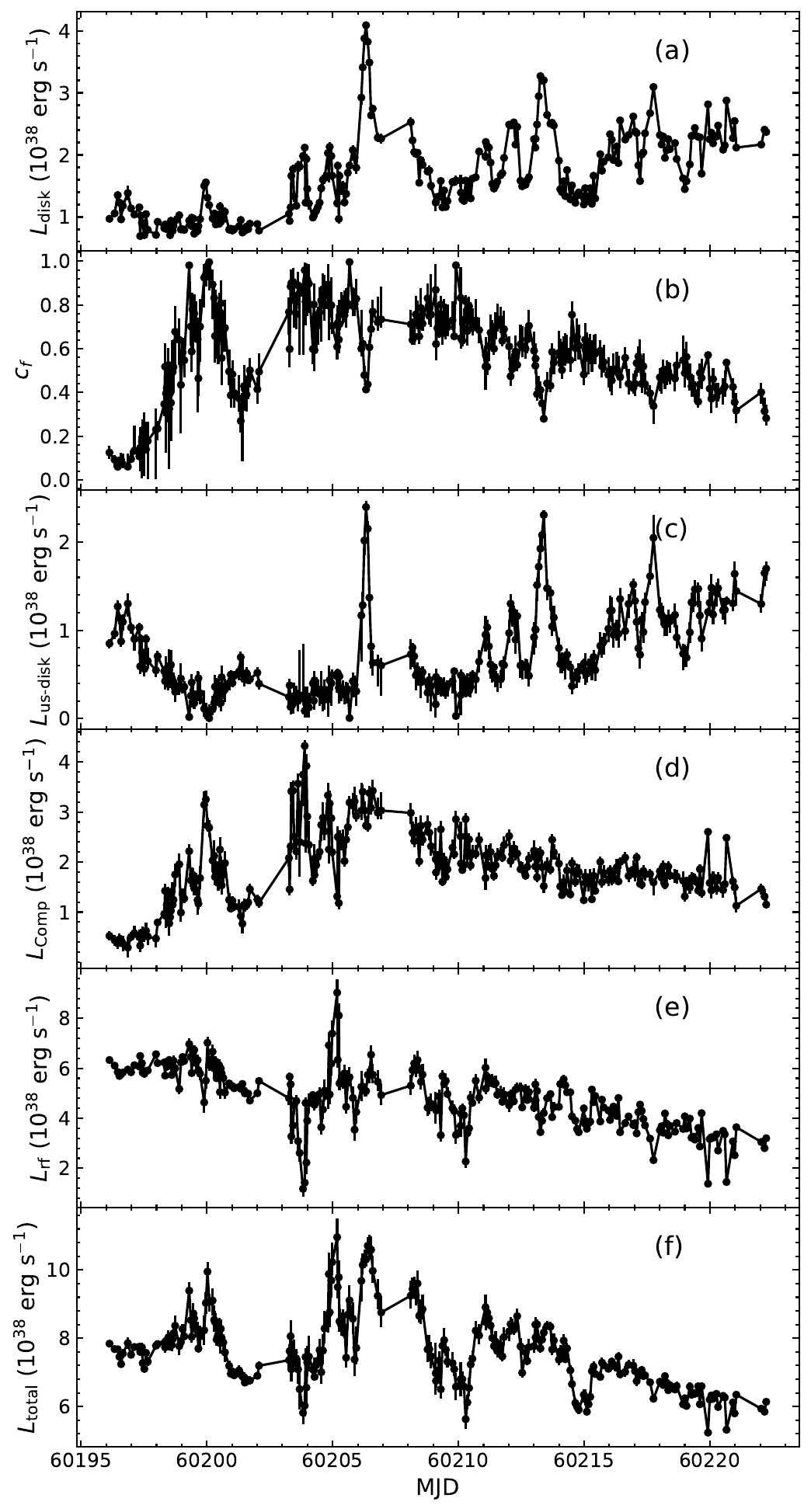}
    \caption{Derived component luminosities in 0.01-1000 keV and the covering fraction $c_f$ based on the spectral model {\tt constant*tbabs*(thcomp$\otimes$diskbb+relxillCp)}. The data points are taken from the spectral fits in \cite{he2025dailyfluctuationspropagatedamply}.} 
    \label{fig:flux_cttdr}
\end{figure}

In Figure~\ref{fig:flux_cttdr}, we show the evolution of the spectral luminosities, which are taken from the spectral fits in \cite{he2025dailyfluctuationspropagatedamply}. 
To better clarify the relationship between the intrinsic disk luminosity and the unscattered disk luminosity, we also show the covering fraction $c_f$ in Figure~\ref{fig:flux_cttdr} (b).
The unscattered disk luminosity $L_{\text{us-disk}}$ directly observed from the accretion disk is derived through $L_{\text{us-disk}} = (1-c_{f})L_{\rm disk}$, and the Compton luminosity is derived through $L_{\rm Comp}=L_{\rm thcomp}-(1-c_f)L_{\rm disk}$. The {\tt relxillCp} luminosity is labeled as $L_{\rm rf}$. The luminosities were computed assuming a distance of $d=3.7\,\text{kpc}$ \citep{Mata2025A&A...693A.129M}.

\section{The correlations for another spectral model}
\label{sec:appendix2}

Following the method described in Section \ref{subsec:spectral analysis}, we performed spectral fitting using the model {\tt constant*tbabs*(diskbb+xillver)}. The {\tt diskbb} model represents the thermal emission directly observed from the disk. The {\tt xillver} model represents the Comptonized emission and reflection emission. Consistent with the previous model configuration, the iron abundance $A_{\rm Fe}$ was fixed at 1.0 and the inclination was set as $40^{\circ}$. The ionization of the accretion disk cannot be well constrained. Considering that \citet{peng2024ApJ...960L..17P} indicates a high ionization, the ionization was fixed at 4.0. The other parameters of {\tt xillver} and the parameters of {\tt diskbb} were left free. In the right column of Figure~\ref{fig:spectral_fitting_all}, we present the unfolded spectra and component decomposition, allowing direct comparison with the left column.

\begin{figure}[tbp]
    \centering
    \includegraphics[width=0.65\textwidth]{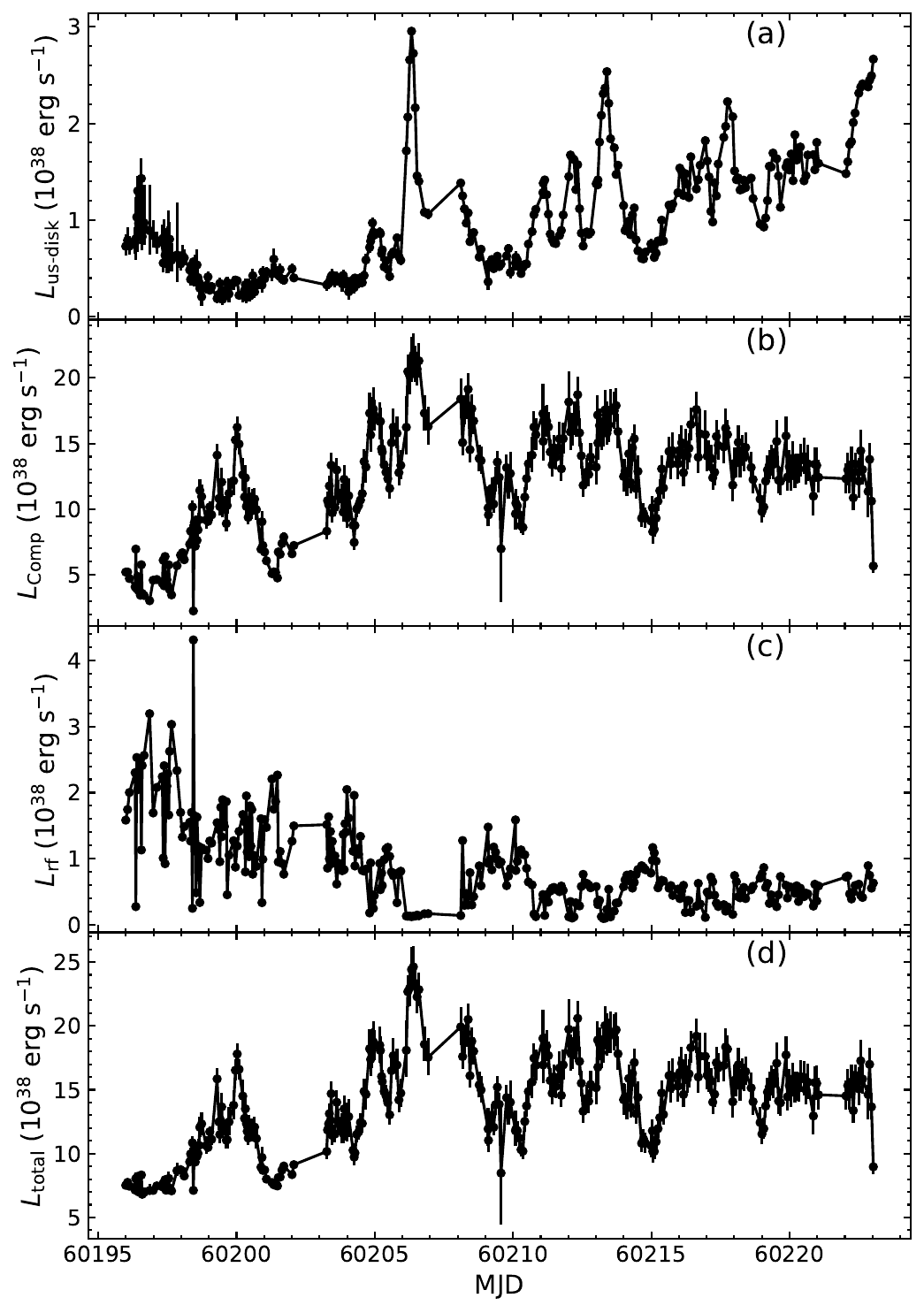}
    \caption{Derived component luminosities in 0.01-1000 keV based on the spectral model {\tt constant*tbabs(diskbb+xillver)}.} 
    \label{fig:flux_ctdx}
\end{figure}

The luminosities in 0.01-1000 keV of disk and Compton components were estimated using {\tt cflux} in {\tt XSPEC}. The {\tt diskbb} luminosity corresponds to the unscattered soft photons which are directly observed. The {\tt xillver} luminosity comprises the luminosity of the Comptonized photons and the reflection photons. By changing the reflection fraction $R_{\rm f}$ to the negative value, the reflection flux can be estimated. The Comptonized luminosity was calculated by $L_{\rm Comp} = L_{\rm xillver} - L_{\rm rf}$. The time evolution of these spectral luminosities is presented in Figure~\ref{fig:flux_ctdx}. It can be seen that, although the absolute luminosities differ from those obtained with model {\tt thcomp$\otimes$diskbb+relxillCp}, the overall trends are consistent between the two spectral models.  

\begin{figure}[tbp]
    \centering 
    \includegraphics[width=0.8\textwidth]{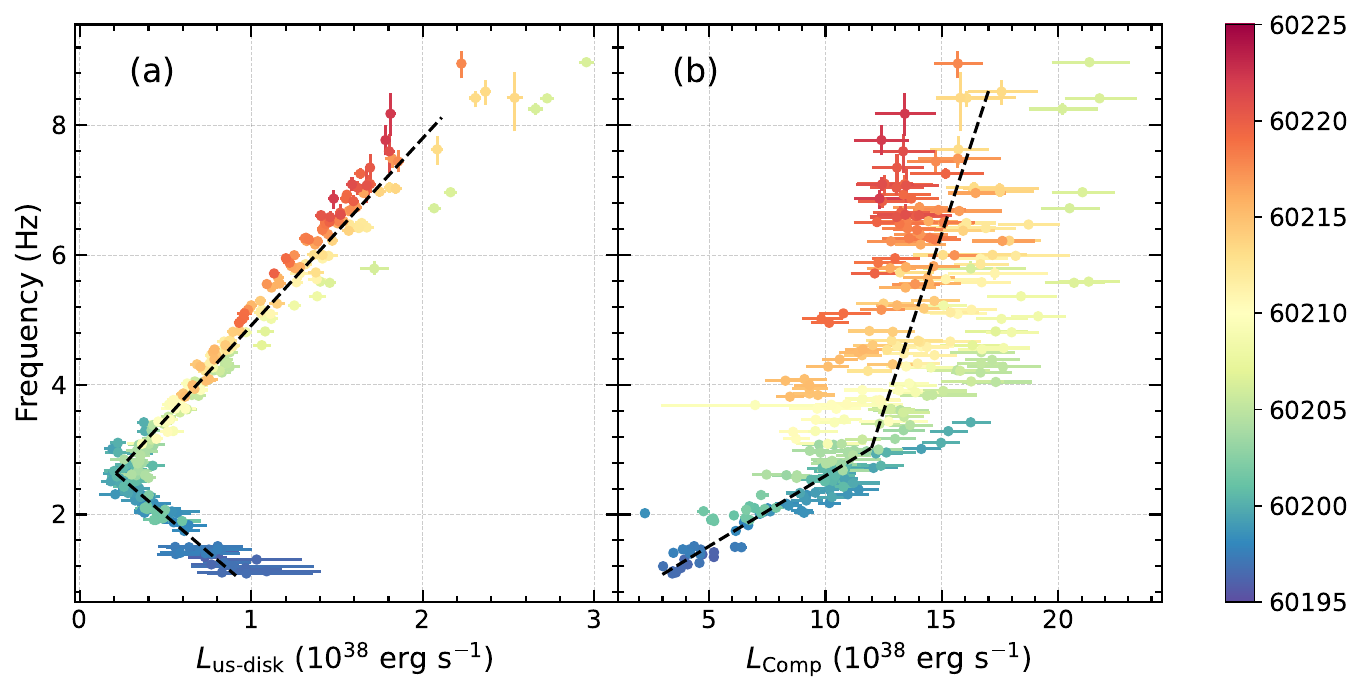}
    \caption{The correlation between QPO frequency and unscattered disk, and Compton luminosity during flare state. The MJD is mapped to the colors of the points. The dashed curves indicate the best-fitting results.} 
    \label{fig:lefre_flux_ctdx}
\end{figure}

In Figure \ref{fig:lefre_flux_ctdx}, we revisit the correlation between QPO frequency and unscattered disk, and Compton luminosity, based on the spectral model {\tt constant*tbabs*(diskbb+xillver)}. Following the methods in Section \ref{subsec:correlations}, we fit these correlations using the function \ref{eq:fit}, and the best-fitting results are plotted with dashed lines in Figure \ref{fig:lefre_flux_ctdx}. The correlation between QPO frequency and unscattered disk luminosity can be fitted with:
\begin{equation} 
\begin{aligned}
    \nu = (-2.26\pm 0.15) L_{\text{us-disk}} (/10^{38} {\rm erg\, s^{-1}}) + (3.11\pm 0.05)&,\\ 
    \text{for}\,\nu < 2.64\, \text{Hz}&;\\
    \nu = (2.89\pm 0.04) L_{\text{us-disk}} (/10^{38} {\rm erg\, s^{-1}}) + (2.03 \pm 0.05)&,\\
    \text{for}\,\nu \geq 2.64\, \text{Hz}&.
\end{aligned}
\end{equation}
The Spearman coefficients for the data below and above $2.64\,\text{Hz}$ are $r_{\rm S}=-0.811$ and $r_{\rm S}=0.981$ with a significance of $4.0 \, \sigma$ and $14.0 \, \sigma$, respectively. For the $\nu$–$L_{\rm Comp}$, the best-fit results are:
\begin{equation} 
\begin{aligned}
    \nu = (0.22\pm 0.02) L_{\rm Comp} (/10^{38} {\rm erg\, s^{-1}}) + (0.42\pm 0.27)&,\\
    \text{for}\,\nu < 3.03\, \text{Hz}&;\\
    \nu = (1.10\pm 0.13) L_{\rm Comp} (/10^{38} {\rm erg\, s^{-1}}) + (-10.09\pm 1.61)&,\\
    \text{for}\,\nu \geq 3.03\, \text{Hz}&.
\end{aligned}
\end{equation}
The Spearman coefficients for the data below and above $3.03\,\text{Hz}$ are $r_{\rm S}=0.880$ and $r_{\rm S}=0.414$ with a significance of $5.6 \, \sigma$ and $2.5 \, \sigma$, respectively. 

The relationships discussed above generally align with the results presented in Section \ref{subsec:correlations}. There are also two-branch correlations with a transition frequency at around $3\,\text{Hz}$. At QPO frequencies below $3\,\text{Hz}$, the QPO frequency $\nu$ is negatively correlated with the unscattered disk emission $L_{\text{us-disk}}$ and positively correlated with the Compton emission $L_{\rm Comp}$. At higher frequencies, the $\nu$–$L_{\text{us-disk}}$ correlation turns positive, while the correlation $\nu$–$L_{\rm Comp}$ remains positive but exhibits a steeper slope and greater scatter. This may further support our speculation in Section \ref{sec: coevolution} that the $\nu$–$L_{\rm Comp}$ relation simply reflects the combined effects of disk emission and the geometry of thin disk and inner hot flow, despite the alternative model lacking explicit parameters to describe this interaction.

\clearpage

\bibliography{sample631.bib}{}
\bibliographystyle{aasjournal}

\end{document}